*Electronic modulation of infrared emissivity in graphene plasmonic resonators*

Victor W. Brar *&, Michelle C. Sherrott *#, Luke A. Sweatlock ~#,
Min Seok Jang *%, Laura Kim*, Mansoo Choi %^ and Harry A. Atwater *

\*   Thomas J. Watson Laboratory of Applied Physics, California Institute of Technology, Pasadena California 91125, United States

~   Nanophotonics and Metamaterials Laboratory, Northrop Grumman Aerospace Systems, Redondo Beach California 90250, United States

&   Kavli Nanoscience Institute, California Institute of Technology, Pasadena, California 91125, United States

#   Resnick Sustainability Institute, California Institute of Technology, Pasadena California 91125, United States

%   Global Frontier Center for Multiscale Energy Systems, Seoul National University, Seoul 151-747, Republic of Korea

^   Division of WCU Multiscale Mechanical Design, School of Mechanical and Aerospace Engineering, Seoul National University, Seoul 151-742, Republic of Korea

**Abstract**

**Electronic control of blackbody emission from graphene plasmonic resonators on a silicon nitride substrate is demonstrated at temperatures up to 250°C. It is shown that the graphene resonators produce antenna-coupled blackbody radiation, manifest as narrow spectral emission peaks in the mid-IR. By continuously varying the nanoresonators carrier density, the frequency and intensity of these spectral features can be modulated via an electrostatic gate. We describe these phenomena as plasmonically enhanced radiative emission originating both from loss channels associated with plasmon decay in the graphene sheet and from vibrational modes in the SiN.**

All matter at finite temperatures emits electromagnetic radiation due to the thermally induced motion of particles and quasiparticles. The emitted spectrum is characterized as:

$$I(\omega, T) = \frac{\hbar \omega^3}{4\pi^3 c^2} \frac{1}{e^{\frac{\hbar \omega}{k_b T}} - 1} \epsilon(\omega, T)$$

where I is the spectral radiant energy density (spectral radiance), T the absolute temperature in Kelvin, $\hbar$ reduced Planck's constant, $\omega$ angular frequency, c the speed of light in vacuum, $k_b$ the Boltzmann constant, and $\epsilon$ the material spectral emissivity. While infrared thermal radiation typically can be assumed to be broadband, incoherent, and isotropic, recent experiments on engineered materials have shown the blackbody emission can be coherent, unidirectional and have narrow spectral features. These structures have included the patterned gratings on metal or silicon carbide surfaces,[1, 2] size-tunable Mie resonances,[3] and frequency selective surfaces.[4] Negative differential thermal emittance has also been explored in materials with strongly temperature dependent emissivity, such as $VO_x$ in the vicinity of its solid state phase transition[5]. In the near-field, where the power of blackbody radiation can exceed the Stefan-Boltzmann limit for far field emission,[6-10] thermal devices have been proposed that display unidirectional flow of heat through control of the blackbody spectrum (i.e. thermal diodes)[11, 12], and that show large amounts of heat transfer between nearby surfaces for solar thermal conversion devices.[13-16] Electronically tunable emissivity has also been demonstrated in the THz regime, where injected charges were used to overdampen a surface phonon polariton mode in a single quantum well.[17]

In this paper, we experimentally demonstrate active electronic control of infrared thermal emission through antenna-mediated modulation of the coupling strength between the thermal emitters and the photonic modes. Our structure is based on field effect tuning of carrier density in graphene plasmonic resonators, which act as antennae to effectively enhance thermal radiative emission within the resonator mode volume. We show that through this mechanism the thermal radiation generated by substrate phonons and inelastic electron scattering in graphene can be enhanced or attenuated and can be fixed within a narrow bandwidth in the mid-IR. The large Purcell factors associated with these plasmonic antennas suggest that this device could potentially control thermal radiation at time scales much faster than the spontaneous emission rate for conventional light emitting diodes and classical blackbody emission sources.

A schematic of our experimental setup is shown in Figure 1a. Our measurements were performed on graphene grown on 25 μm thick copper foils using established chemical vapor deposition growth techniques.[18, 19] The graphene was transferred to a 1μm thick low stress silicon nitride ($SiN_x$) membrane with 200nm of Au deposited on the opposite side that is used as both a reflector and a backgate electrode. Nanoresonators with widths ranging from 20-70nm were then patterned over 60×60μm$^2$ areas into the graphene using 100keV electron beam lithography (see Methods). A typical gate-dependent resistance curve for one of our structures is shown in Figure 1. The peak in resistance corresponds to the charge neutral point (CNP) of the graphene, where the Fermi level is aligned with the Dirac point and the carrier density is minimized. After the CNP for each structure was measured, a capacitor model[20] was used to determine the carrier density corresponding to each applied gate voltage (see Supporting Information).

The device geometry described above was previously used as a gate-tunable absorber in the mid-IR, where a large enhancement in absorption was observed when the graphene plasmonic resonance was matched to the same energy as the $\lambda/4n_{SiN}$ resonance condition in the 1μm $SiN_x$ layer, which occurred at 1400cm$^{-1}$.[21] In those experiments it was shown that the total absorption in the graphene nanoresonators could be tuned from 0 to up to 24.5% for large carrier densities, and up to ~10% for the carrier densities used in this work, where the maximum applied field is limited by Poole-Frenkel tunneling in the $SiN_x$ (See Supporting Information).[21, 22] For blackbody emission measurements, the device was connected to a temperature-controlled stage consisting of a 100μm thick layer of sapphire on 2mm copper on a heated silver block that can vary in temperature from room temperature to 250°C. The device and stage were held at a pressure of 1-2 mTorr during emission measurements. Gate-dependent emission spectra were measured using a Fourier transform infrared (FTIR) microscope operating such that emitted light from the heated device passes through a KBr window and is collected in a Cassegrain objective, collimated and passed through the interferometer in the FTIR before being focused on a liquid nitrogen-cooled HgCdTe detector. For polarization dependent measurements a wire grid polarizer was placed in the collimated beam path. As a reference a SiN/Au membrane was coated with an optically thick layer of black soot deposited using a candle. Soot is known to be a thermal emitter that approximates an ideal blackbody with emissivity approaching unity across the mid-IR.[5]

Figure 2 (left axis) shows the emitted radiation at 250°C from the black soot reference, a bare $SiN_x$/Au membrane, and from a 40nm graphene nanoresonator array at 250°C under doped ($4.9 \times 10^{12}$ / $cm^2$) and undoped conditions. On the right axis of Fig. 2 we plot the change in emissivity corresponding to the observed change in emitted light from the undoped to doped graphene resonators. This change in emissivity is calculated assuming unity emissivity at all frequencies for the black soot reference and normalizing accordingly. As can be seen in the figure, increasing the carrier density of the graphene nanoresonators leads to increases in emissivity near 750$cm^{-1}$ and 1400$cm^{-1}$.

In order to explore these gate-tunable emissivity features further, we investigate their polarization dependence (Fig. 3(c)), as well as their behavior as the nanoresonator doping and width is varied, as shown in Fig 3 (a,b). These results indicate that the intensity, width and energetic position of the thermal radiation feature near 1400$cm^{-1}$ is strongly polarization dependent and is widely tunable. The energy of this feature increases as the nanoresonator width is decreased and as the carrier density is increased, while the intensity of this feature increases with carrier density, and is largest in 40nm resonators, when it occurs closest to the $\lambda/4n_{SiN}$ resonance condition of the SiN at 1400$cm^{-1}$. Because Kirchoff's Law dictates that thermal emissivity is equal to absorptivity, these observations are consistent with previously reported absorption measurements performed on identical samples that showed a narrow absorption feature near 1400$cm^{-1}$.[21] The lower energy emissivity modulation feature near 750$cm^{-1}$ shows different behavior than the higher energy peak. Namely, the low energy feature shows an extremely weak polarization dependence, and also shows no noticeable dependence on graphene nanoresonator width. As the carrier density is increased, there is an increase in intensity for this feature, but it shows no spectral shift. Finally, unlike the higher energy peak, the lower energy peak is also observed in the bare, unpatterned graphene, where it appears as a slightly narrower feature. The absorption properties of this device near the energy range of the lower energy feature was not discussed in previously reported work due to the low energy cutoff of the detector used in that work.

We explain the above phenomena as electronic control of thermal radiation due to a combination of plasmon-phonon and plasmon-electron interactions, Pauli-blocking effects, and non-radiative transfer processes between the $SiN_x$ and the graphene sheet. While Kirchoff's law dictates that the thermal equilibrium emissivity must be equal to the absorbtivity for any material, the precise, microscopic mechanism of thermal emission is interesting when the system includes highly confined optical modes, as is the case here. We now describe in detail the interplay of these microscopic processes for both the high energy and low energy features.

We first explain the prominent feature at 1400cm[-1] as being due to a Fabry-Perot type plasmonic resonance from the patterned graphene. The width and doping dependence of the 1400cm[-1] feature follows the behavior expected for graphene plasmonic modes, and is consistent with reflection measurements.[21] Specifically, the graphene plasmon resonant frequency should vary as $\omega_p \propto n^{1/4} W^{-1/2}$ and this behavior is reflected in the emission spectra in which we observe a blue-shift of the plasmonic resonance as we increase doping and decrease the width of the graphene nanoresonators. Furthermore, the intensity of this higher energy feature increases with graphene carrier density, an effect that results from the increased polarizability of the resonant plasmonic modes. Finally, this feature is strongly polarization dependent - as we would expect laterally bound graphene plasmonic resonances to be - and vanishes quickly as we transition from probing radiation 90° to 0° relative to the nanoresonator axis.

In order to understand the source of the thermally excited plasmons in the graphene nanoresonators, we note that the microscopic processes that lead to plasmonic loss in graphene should become plasmon-generating processes when the sample is heated. For the case of the 1400cm[-1] feature we observe here, the plasmon decay (and therefore plasmon generating) processes are mediated by the same pathways that limit the electron mobility of the graphene, such as defect scattering, impurity scattering, and inelastic electron-electron and electron-phonon interactions.[23-27] Additionally, plasmons have been shown to decay via loss channels associated with the edges of graphene nanostructures, and by coupling to substrate phonons.[23, 27] For a bare graphene sheet, the plasmons generated by thermal emission do not couple well to free space and are thus non-radiative. Upon patterning the graphene, however, the plasmonic resonances can effectively out-couple radiation, and the plasmon decay processes become free-space thermal emission sources by exciting resonant plasmonic modes which then radiate.

The resonant enhancement of emission from plasmon generating processes is in competition with the blocking of interband transitions which act as thermal emitters in the undoped graphene, but are forbidden due to Pauli blocking when the sheet is doped.[28, 29] The role of interband transitions can be seen most clearly in the bare graphene emissivity spectra in Fig. 3b where there is a broad decrease in emissivity near 1400cm$^{-1}$ at higher carrier densities. While interband transitions should occur across a wide range of frequencies, in the backreflector geometry we use here, thermal emission from the surface can either constructively or destructively interfere with itself and is thus most prominent at 1400cm$^{-1}$, the $\lambda/4n_{SiN}$ frequency of the SiN$_x$ layer. For patterned graphene areas, however, we find that doping the graphene allows for the resonant plasmonic modes to create and emission enhancement that outweighs the decrease in emission due Pauli blocking, and thus we get a net increase in emission near 1400cm$^{-1}$.

As mentioned above, in addition to out-coupling of radiation due to plasmon loss mechanisms in the graphene, the plasmonic resonators also interact with vibrations in the SiN$_x$ substrate. When the SiN$_x$ is heated, the plasmonic modes act as antennae to enhance the spontaneous thermal radiation from the nearby SiN$_x$. The enhancement of the spontaneous emission radiative rate and of the quantum efficiency arising from dipole emitters' proximity to a dipole optical antenna is well known,[30-32] and is attributed to increasing the probability of radiation by modification of the photonic mode density.[33] The rate enhancement is correlated to the strong polarizability of the graphene at its plasmonic resonance which enhances the outcoupling of thermal radiation from the SiN$_x$. In particular, the radiative rate is expected to be most strongly amplified in the top 10nm of the SiN$_x$ in accordance with the approximate effective mode volume of the resonant graphene plasmon. However, we also expect that thermal emitters in the SiN$_x$ near the graphene surface should experience non-radiative decay which competes with this enhancement effect.[34] We therefore assign the net increase of thermal emission as a combination of the confined plasmonic modes out-coupling energy from thermal excitations in the graphene as well as thermal phonons in the SiN$_x$. These processes exceed the decrease in emission associated with non-radiative quenching effects of the graphene on the nearby SiN$_x$, as well as the blocking of interband transitions in the graphene sheet.

We next consider the feature at 730cm$^{-1}$ which is located at the energy of a strongly absorbing phonon in the SiN$_x$ that creates an emission peak at raised temperatures, as observed in

Fig. 2. This emission peak is influenced in a number of ways by the presence of the graphene. First we note that, as shown in Fig. 2, emission from 720 to 1100cm$^{-1}$ is decreased for both doped and undoped graphene ribbons on SiN$_x$ in comparison to the bare membrane. We attribute this to non-radiative energy transfer processes from the SiN$_x$ thermal excitations to the graphene sheet.[34] For undoped graphene, these processes are represented by interband transitions in the graphene sheet that have been predicted and shown to dramatically quench the emission of nearby dye molecules.[35] As the graphene becomes doped, interband transitions are blocked, but new non-radiative pathways are introduced in the form of propagating plasmons in the graphene sheet.[34]

The SiN$_x$ phonon at 730cm$^{-1}$ can also couple to the plasmons in the graphene to create a new surface phonon plasmon polariton (SPPP) mode, similar to what has been observed for graphene on SiO$_2$ and h-BN.[23, 27, 36] Similar to the resonant plasmonic modes described earlier, this mode can also enhance the thermal emission into free space, and this emission should increase with the carrier density of the graphene sheet. In order to calculate the possible contribution of this mode we performed full-wave finite element electromagnetic simulations to calculate the full plasmonic bandstructure for the graphene/SiN$_x$ system, as shown in Fig. 6. This figure shows, indeed, that the graphene plasmon spectrum has been perturbed by the SiN$_x$ phonons, and that a new SPPP mode exists as a flat band is introduced near 650cm$^{-1}$ along with a fainter, also flat band at 750cm$^{-1}$. While the lower branch of SPPP mode is expected to show gate tunable effects on the graphene nanoribbon emissivity, it should also show a strong polarization dependence, which is not observed in Fig 3c. Additionally, this low energy feature is observed in the emissivity modulation of the bare graphene sheet where the SPPP mode should show weak out-coupling behavior. In contrast, the weaker phonon branch near 750cm$^{-1}$ crosses the lightline, and thus does not require patterning to couple to freespace, and should not display an intensity dependent polarization dependence. Due to these observations, we determine that the feature near 730cm$^{-1}$ is created by both non-radiative processes between SiN$_x$ phonons and the nearby graphene sheet, as well as by the weaker branch of the graphene/SiN$_x$ SPPP mode.

To better understand and quantify our emission features from the graphene-SiN$_x$ interactions, we used a finite element method to calculate the electromagnetic power density ($\nabla \cdot \vec{S}$) associated with the absorption of plane waves incident on 40nm graphene nanoresonator on a $SiN_x$/Au substrate at $4.9\times10^{12}cm^{-2}$ carrier density using parameters described in our previous works.[21] The electromagnetic power density models where power is absorbed, and therefore also indicates where far field thermal emission originates. The results of these simulations are shown in Fig. 4(a) at $1413cm^{-1}$, corresponding to the graphene plasmon mode. We observe a strong enhancement of the density of electromagnetic power absorption near the graphene resonators. On this resonance, there is a significant amount of power being absorbed into the graphene; however, it can clearly be seen that in the region in which the graphene plasmon extends into the $SiN_x$, there is an enhancement of power absorption, which would translate into an increased rate of spontaneous emission from this part of the substrate. To further quantify this, we integrate the power densities over each material for undoped and doped graphene and see that, the power absorbed into the top 10nm of the $SiN_x$ increases with the increased graphene nanoresonator doping as shown in Fig. 4(b). At $\sim1400cm^{-1}$, it is observed that there is weak power absorption in the top layer of $SiN_x$ for undoped graphene, and we see only the interband transitions contributing in the graphene itself. Then as the doping is increased to $E_F=0.25eV$, the graphene plasmon can be excited and so absorption in the graphene and top 10nm of $SiN_x$ increases due to the effects described above. For comparison we show the absorption features from the remaining bottom 990nm of $SiN_x$. It is important to note that this finite element model does not account for the non-radiative processes discussed above. This model only indicates how graphene plasmons interact with a homogenous, lossy medium and not for the way that localized dipole moments interact with the graphene sheet which is the origin of the non-radiative quenching effects.

In order to quantify the thermally radiated power of this structure, we consider Planck's law for spectral radiance using the black soot as a reference with $\varepsilon=1$, and including our 50x50 $\mu m^2$ collection area and the 1.51 steradians covered by the 0.65 NA objective. We plot these results for different temperatures in Fig. 5, showing an increase in the thermally radiated power that is modulated by the graphene sheet, and a maximum thermal power modulation of $200pW/cm^{-1}$ at $1400cm^{-1}$ (7.1 μm). These calculations indicate that a 1x1 $mm^2$ device patterned with 40nm resonators held at $250°C$ could act as an electronically controllable mid-IR source that would emit 8μW over 100 $cm^{-1}$ of bandwidth. This compares favorably to commercial mid-IR LEDs at 7μm, which emit 2μW over similar bandwidths.[37] We also note that the maximum temperature and gate bias applied in these experiments was not limited by the graphene but by the $SiN_x$ dielectric, which is known to exhibit Poole-Frenkel tunneling at high temperatures.[22] By choosing a dielectric that can withstand higher temperatures, such as $SiO_2$, larger powers could be achieved in such devices.

In conclusion, we have demonstrated the direct electronic control of Mid-IR thermal radiation using graphene plasmonic nanoresonators. We show that the graphene plasmonic modes can act to enhace the thermal radiation from the $SiN_x$ membrane as well as excitations in the graphene sheet. We have developed a structure with tunable narrowband emission at a range of frequencies in the mid-IR due to graphene nanostructure resonances, and we have shown that this emission can be changed statically with resonator dimensions, and actively with charge carrier density via the application of a gate bias. We estimate that the power emitted from this structure with a 1mm$^2$ areal coverage could exceed that of mid-IR LEDs.

Acknowledgements:
This work was supported by the Department of Energy Office of Science, Basic Energy Sciences under Contract No. DE-FG02-07ER46405 (M.C.S.,V.W.B. and H.A.A.). M. S. J. and M. C. acknowledge support from the Global Frontier R&D Program on Center for Multiscale Energy Systems funded by the National Research Foundation under the Ministry of Science, ITC & Future Planning, Korea (2011-0031561, 2011-0031577). M.C.S. acknowledges support from a Resnick Institute Graduate Fellowship. M.S. J. acknowledges a post-doctoral fellowship from the POSCO TJ Park Foundation. V.W.B. acknowledges support from a Kavli Nanoscience Postdoctoral Fellowship, and use of facilties of the Kavli Nanoscience Institute.

**Figures**

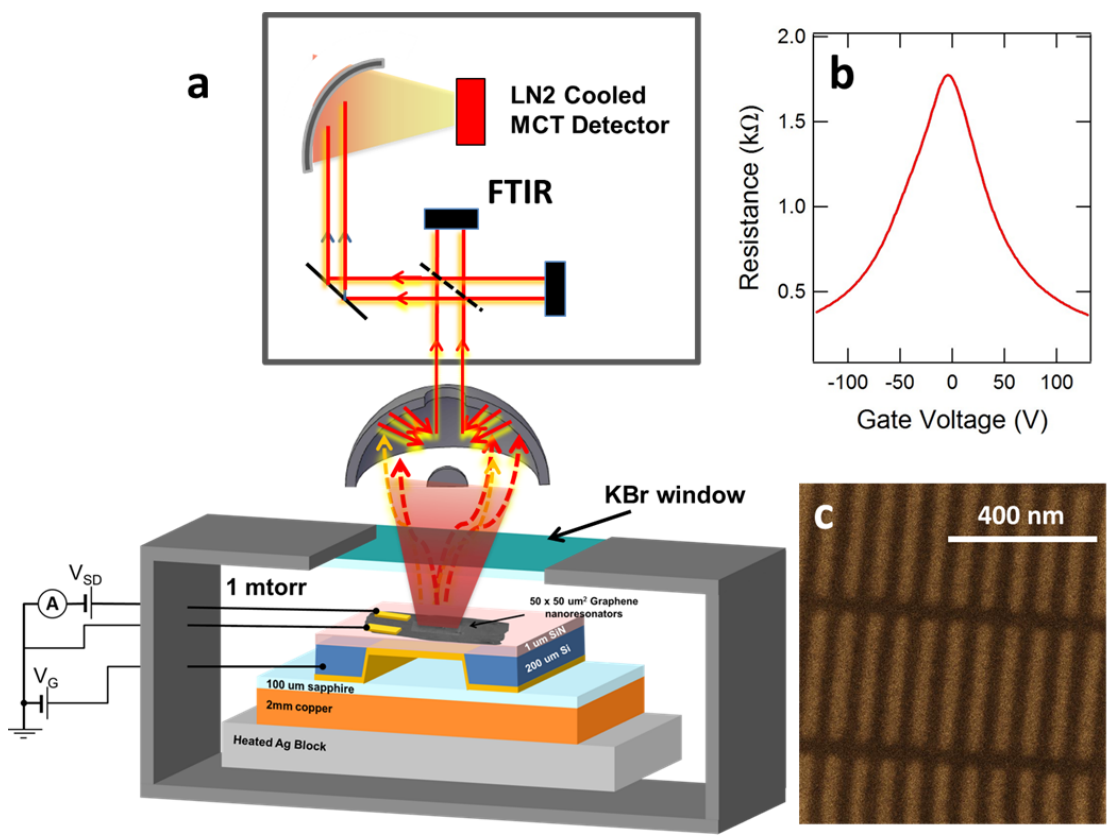

**Figure 1:** Schematic of experimental setup. **(a)** Graphene structure on temperature-controlled stage with FTIR emission measurement configuration. Graphene structure consists of 80 × 80 μm$^2$ nanoresonator arrays on a 1μm thick SiN$_x$ membrane with 200nm Au backreflector. The graphene was grounded through Au(100nm)/Cr(3nm) electrodes that also served as source-drain contacts. A gate bias was applied through the SiN$_x$ membrane between the underlying Si frame and graphene sheet. Temperature controlled stage consists of 100μm thick sapphire on 2mm Cu on an Ag block. Emission measurements were taken at different temperatures via FTIR using a LN$_2$ cooled MCT detector.
**(b)** A resistance vs gate voltage curve of the graphene sheet showing a peak in the resistance at the charge neutral point (CNP), when the Fermi level (E$_F$) is aligned with the Dirac point.
**(c)** A representative SEM image of 30nm graphene nanoresonators.

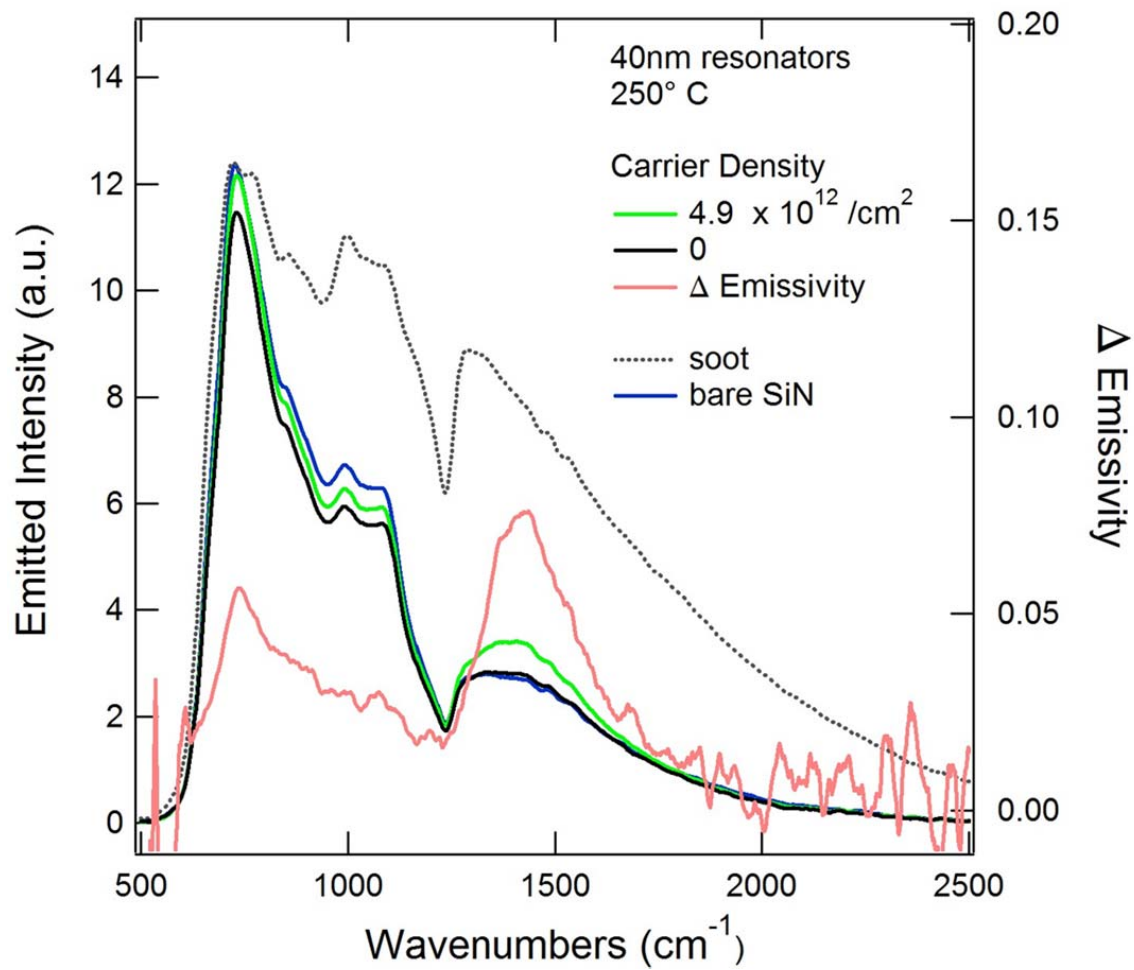

**Figure 2:** Normalization scheme adopted for this experiment including reference soot emission spectrum (left axis) taken to have emissivity of unity. Emitted intensity at a given temperature for bare SiN$_x$, graphene at charge neutral point (CNP) and increased carrier density (left axis). Change in emissivity of structure from CNP to doped graphene, normalized to emission spectrum of soot (right axis). Enhancement of emissivity is observed due to increased charge carrier density in graphene.

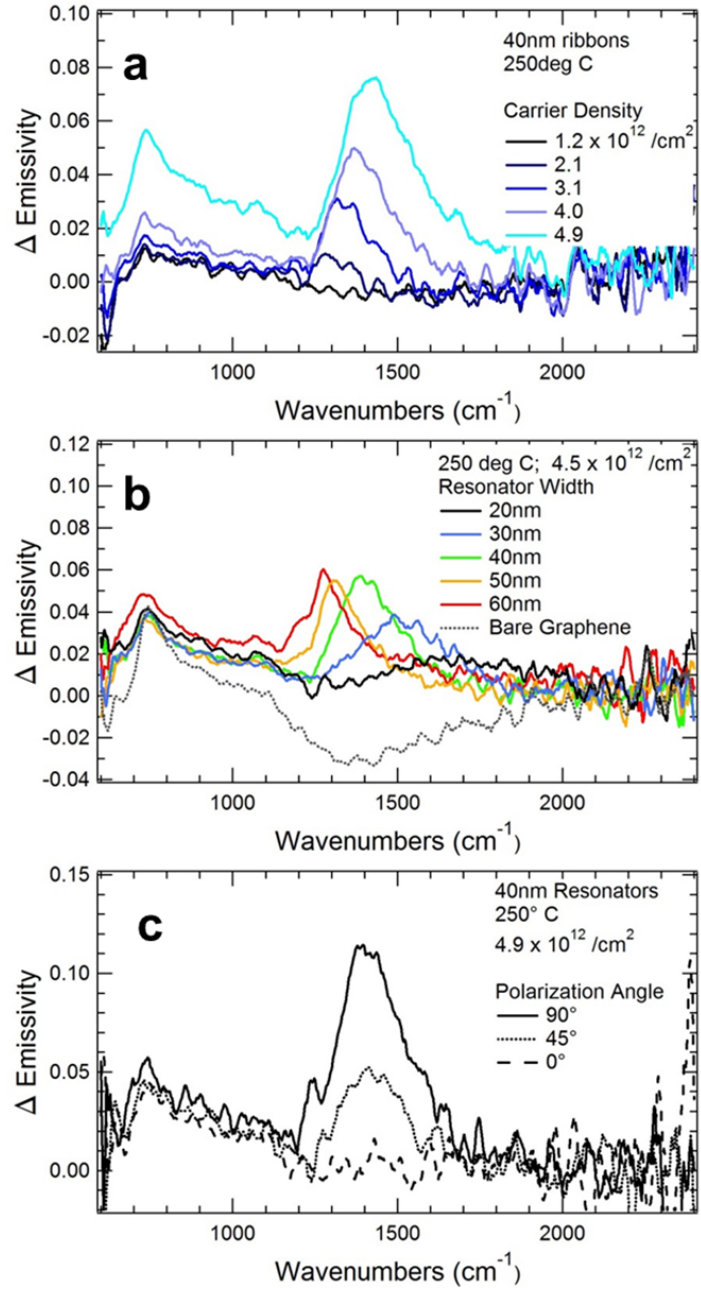

**Figure 3:** Modulation of emissivity and thermal emission. Emissivity calculated using a unity emissivity soot reference at the same temperature. **(a)** Carrier density dependence of emissivity modulation for a fixed temperature and nanoresonator width. **(b)** Emissivity modulation for different nanoresonator geometries as well as unpatterned graphene at a fixed temperature and carrier density. **(c)** Emissivity change for different polarizations of light for a fixed temperature, resonator width, and carrier density.

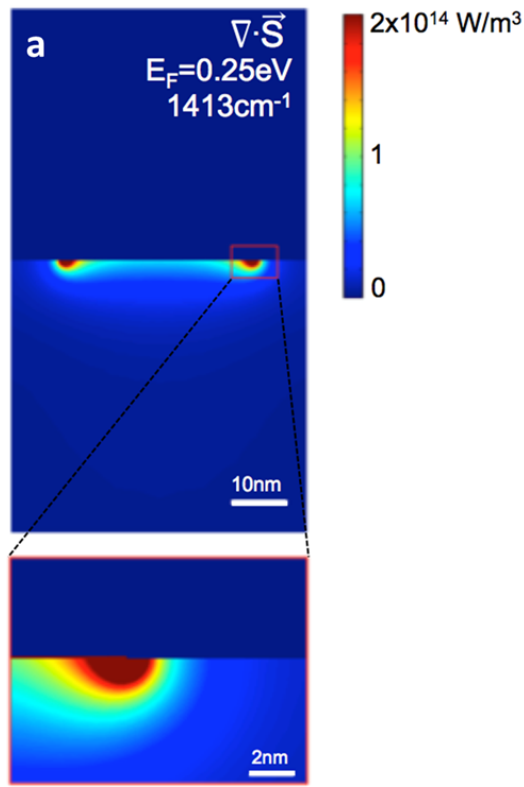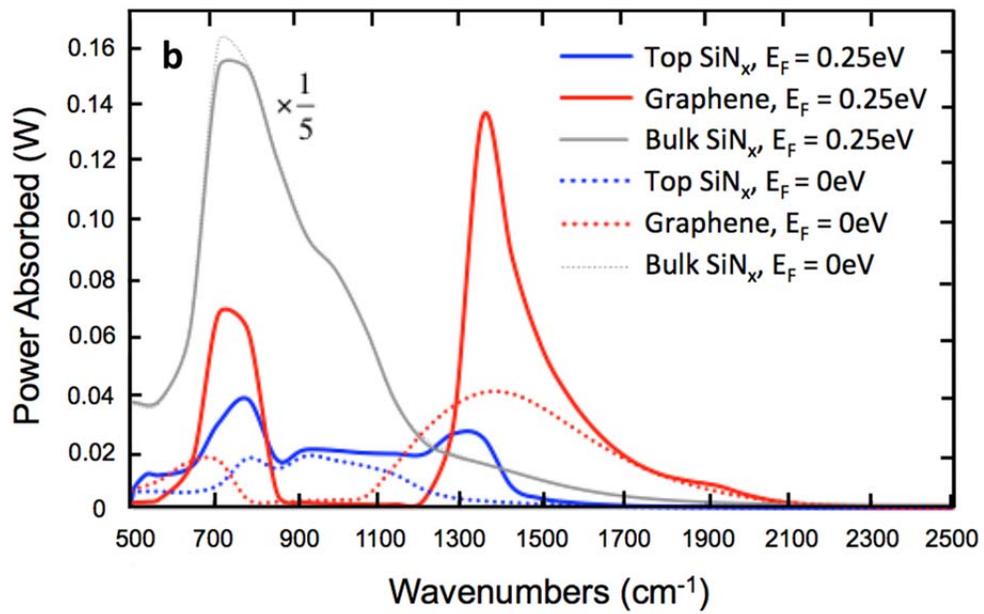15

**Figure 4: (a)** Calculated 2D plot of $\nabla \cdot \vec{S}$ electromagnetic power density in graphene/SiN$_x$ structure with vacuum above obtained from finite element electromagnetic simulation. Plotted at 1413cm$^{-1}$ (graphene plasmon peak) at E$_F$=0.25eV. Enhancement of power density noted closest to graphene surface then decaying into SiN$_x$ substrate.
**(b)** Integrated power density in 40nm width graphene resonator, the top 10nm of SiN$_x$ (Top SiN$_x$), and the remaining 990nm of SiN$_x$ (Bulk SiN$_x$) at E$_F$ = 0eV and E$_F$ = 0.25eV.

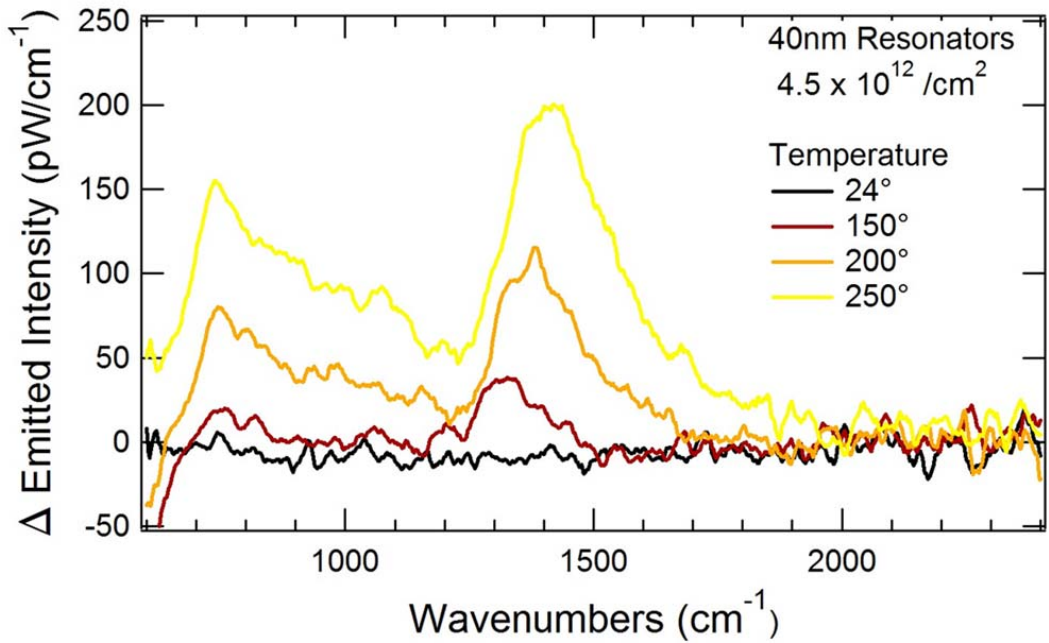

**Figure 5:** Thermally radiated power from Graphene/SiN$_x$/Au structure at varying temperatures for a nanoresonator width of 40nm and carrier density of 4.9x10$^{12}$/cm$^2$. Calculated using black soot reference, based on a 0.65 NA objective and a 50x50μm$^2$ collection area. A maximum modulation of 200pW/cm$^{-1}$ is calculated.

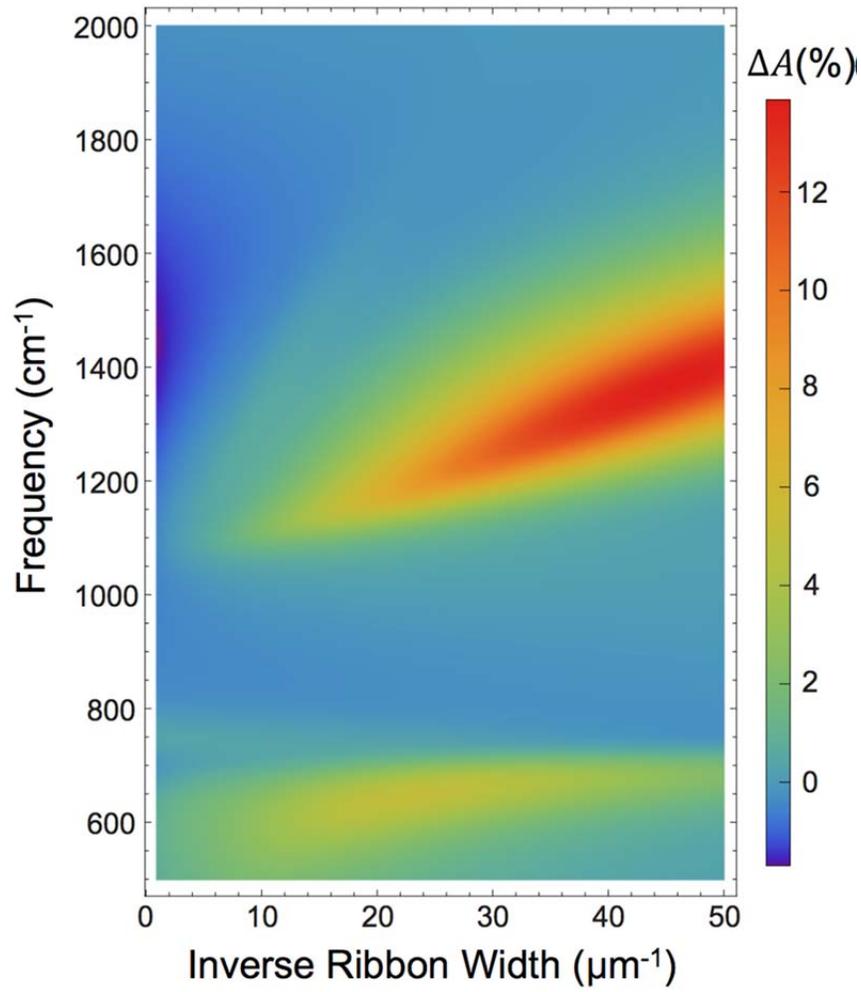

**Figure 6:** Theoretical change in absorption (Δ*A*) as a function of inverse ribbon width at $E_F$=0.25eV. Numerical full-field electromagnetic simulation has been performed using a finite element method under the assumption of normal light incidence.